\documentclass[
  ,draft            
  ,numberedheadings 
  ]
  {aipproc}

\usepackage[mathscr]{eucal}
 \usepackage{amsmath, amssymb}

\def\dmf{\dot{\mathfrak{M}}}
 \def\msE{\mathscr{E}}
  \def\sun{\hbox{$\odot$}}
   
\newcommand{\be}{\begin{equation}}
\newcommand{\ee}{\end{equation}}
\newcommand{\bdm}{\begin{displaymath}}
\newcommand{\edm}{\end{displaymath}}
\def\ga{\mathrel{\mathchoice {\vcenter{\offinterlineskip\halign{\hfil
$\displaystyle##$\hfil\cr>\cr\sim\cr}}}
{\vcenter{\offinterlineskip\halign{\hfil$\textstyle##$\hfil\cr
>\cr\sim\cr}}}
{\vcenter{\offinterlineskip\halign{\hfil$\scriptstyle##$\hfil\cr
>\cr\sim\cr}}}
{\vcenter{\offinterlineskip\halign{\hfil$\scriptscriptstyle##$\hfil\cr
>\cr\sim\cr}}}}}
\layoutstyle{6x9}

\begin{document}

\vspace{-5.5cm}
\noindent {\it Published in Astrophysics and Space Science, April 2013}

\vspace{1cm}

\title{A new look at the origin of the 6.67\,hr period X-ray pulsar 1E~161348-5055}


\author{N.R.\,Ikhsanov}{
  address={Pulkovo Observatory, Pulkovskoe Shosse 65, Saint-Petersburg 196140, Russia}
}

\author{V.Y.\,Kim}{
  address={Pulkovo Observatory, Pulkovskoe Shosse 65, Saint-Petersburg 196140, Russia}
}

\author{N.G.\,Beskrovnaya}{
  address={Pulkovo Observatory, Pulkovskoe Shosse 65, Saint-Petersburg 196140, Russia}
}

\author{L.A.\,Pustil'nik}{
  address={Israel Cosmic Ray and Space Weather Center, Tel Aviv University, P.O. Box 39040, Tel Aviv, 69978, Israel}
}

\begin{abstract}
The point X-ray source 1E~161348-5055 is observed to display pulsations with the period 6.67\,hr and $|\dot{P}| \leq 1.6 \times 10^{-9}\,{\rm s\,s^{-1}}$. It is associated with the supernova remnant RCW\,103 and is widely believed to be a $\sim 2000$\,yr old neutron star. Observations give no evidence for the star to be a member of a binary system. Nevertheless, it resembles an accretion-powered pulsar with the magnetospheric radius $\sim 3000$\,km and the mass-accretion rate  $\sim 10^{14}\,{\rm g\,s^{-1}}$. This situation could be described in terms of accretion from a (residual) fossil disk established from the material falling back towards the star after its birth. However, current fall-back accretion scenarios encounter major difficulties
explaining an extremely long spin period of the young neutron star. We show that the problems can be avoided if the accreting material is magnetized. The star in this case is surrounded by a fossil magnetic slab in which the material is confined by the magnetic field of the accretion flow itself. We find that the surface magnetic field of the neutron star within this scenario is $\sim 10^{12}$\,G and that a presence of $\ga 10^{-7}\,{\rm M_{\sun}}$ magnetic slab would be sufficient to explain the origin and
current state of the pulsar.
\end{abstract}

\maketitle


  \section{Introduction}

The point X-ray source 1E~161348-5055 (hereafter 1E1613) is observed to display pulsations with the period $P_* \simeq 6.67$\,hr  \citep{De-Luca-etal-2006, Esposito-etal-2011}. It is located near the center of the supernova remnant RCW\,103 \citep{Tuohy-Garmire-1980} of the age $\tau_* \sim 2000$\,yr \citep{Nugent-etal-1984, Carter-etal-1997} situated at the
distance of 3.3\,kpc \citep{Caswell-etal-1975, Reynoso-etal-2004}. The X-ray luminosity of the pulsar varies in the interval $L_{\rm X} \simeq 10^{33}-10^{35}\,{\rm erg\,s^{-1}}$ on a timescale of a few years. The X-ray emission comes from a local ($a_{\rm p} \sim 600$\,m) region heated up to a temperature $kT \sim 0.6-0.8$\,keV \citep{Gotthelf-etal-1997, Gotthelf-etal-1999, De-Luca-etal-2006, Esposito-etal-2011}. The upper limit on derivative of the period of pulsations $|\dot{P}| \leq 1.6 \times 10^{-9}\,{\rm s\,s^{-1}}$ has recently been reported by \citet{Esposito-etal-2011}.

It is widely adopted that 1E1613 is a young ($\sim 2000$\,yr) neutron star. In X-rays it resembles an accretion-powered pulsar \citep{Gotthelf-etal-1999} which accretes material at the rate \citep[see e.g.][]{Lamb-etal-1973}
 \be\label{dmf0}
\dmf_* \simeq 5 \times 10^{13}\ m^{-1}\ R_6\ L_{34}\ \ {\rm g\,s^{-1}}
 \ee
and is surrounded by the magnetosphere of the radius
 \be\label{r0}
 r_{\rm mag} \simeq 3 \times 10^8\ R_6^3\ \left(\frac{a_{\rm p}}{600\,{\rm m}}\right)^{-2}\ {\rm cm}.
 \ee
Here $R_6 = R_{\rm ns}/10^6$\,cm and $m = M_{\rm ns}/1.4\,M_{\sun}$ are the radius and mass of the neutron star, and $L_{34}$ is the average X-ray luminosity of the pulsar in units of $10^{34}\,{\rm erg\,s^{-1}}$.

Observations give no evidence for 1E1613 to be a close binary system \citep{Wang-etal-2007}. Nevertheless, \citet{Li-2007} have argued that the source of the accreting material can be a fossil (residual) disk formed by the supernova ejecta fall-back. The lifetime of a fall-back disk is about $10^4-10^5$\,yr and during this time it can supply enough material to power the observed luminosity \citep{Chatterjee-etal-2000}.

While the conventional accretion scenario provides us with the simplest interpretation of the emission spectrum, it encounters major difficulties explaining the pulsar spin evolution. The observed spin-down behavior indicates that the spin period of the pulsar is smaller than (or comparable to) a so called equilibrium period, $P_{\rm eq}$, which is defined by equating the spin-up and spin-down torques applied to the star from the accreting material. If 1E1613 were a regular neutron star accreting from a Keplerian disk its
equilibrium period would be as short as \citep{Alpar-2001, Ikhsanov-2007}
 \be
P_{\rm eq}^{\rm (Kd)} \sim 30\,\mu_{30}^{6/7}\,\dmf_{14}^{-3/7}\,m^{-5/7}\ {\rm s}.
 \ee
Here $\mu_{30}$ is the dipole magnetic moment of the neutron star in units of $10^{30}\,{\rm G\,cm^3}$ and $\dmf_{14} =\dmf_*/10^{14}\,{\rm g\,s^{-1}}$. Thus, to describe the pulsar spin evolution one has to assume that either its surface field in the current epoch is in excess of $5 \times 10^{15}$\,G or the accretion picture differs from the Keplerian disk. The first possibility has been already discussed by \citet{De-Luca-etal-2006, Li-2007} and \citet{Pizzolato-etal-2008}. Here we address analysis of the second possibility. We show that the star under certain conditions can be surrounded by a fossil magnetic slab (Sect.\,\ref{geom}). Using the spin-down torque applied to the star from the slab (Sect.\,\ref{torque}) we find that the current state of the pulsar (Sect.\,\ref{current}) as well as its previous spin evolution (Sect.\,\ref{history}) can be explained provided the surface magnetic
field of the star is $\sim 10^{12}$\,G. Our results are briefly discussed in Sect.\,\ref{conclusions}.

 \section{Geometry of the Fall-back accretion flow}\label{geom}

We consider a fall-back accretion \citep{Michel-1988, Woosley-Chevalier-1988} onto a magnetized neutron star. According to this
scenario the  star of the mass $M_{\rm ns}$ and the dipole magnetic moment $\mu$ after its birth is embedded into a dense gaseous medium of the density $\rho_{\infty}$. As the star moves though the gas with a relative velocity $v_{\rm rel}$ it captures material at a rate $\dmf = \pi r_{\rm G}^2 \rho_{\infty} v_{\rm rel}$, where $r_{\rm G} = 2GM_{\rm ns}/v_{\rm rel}^2$ is the Bondi radius.

The material inside Bondi radius initially follows ballistic trajectories moving towards the star with the free-fall velocity, $v_{\rm ff}(r) = (2GM_{\rm ns}/r)^{1/2}$. If it is non-magnetized and does not possess an angular momentum the accretion occurs in a spherically symmetrical fashion on a dynamical (free-fall) timescale, $t_{\rm ff} = r/v_{\rm ff}$. The flow in this case is decelerated by the stellar magnetic field at the Alfv\'en radius, $r_{\rm A} = (\mu^2/\dmf \sqrt{2GM_{\rm ns}})^{2/7}$, penetrates into the magnetosphere and reaches the stellar surface flowing along the field lines \citep{Lamb-etal-1977}.

A different geometry of the accretion flow can be expected if the captured material is frozen into a large-scale magnetic field, $B_{\rm f}$. In this case the magnetic pressure, $\msE_{\rm m}(r) = B_{\rm f}^2(r)/8 \pi \propto r^{-4}$, in the spherical flow increases rapidly and reaches the ram pressure, $\msE_{\rm ram}(r) = \rho(r) v_{\rm ff}^2(r) \propto r^{-5/2}$, at a distance \citep{Shvartsman-1971}
 \be\label{rsh}
 R_{\rm sh} = \beta_0^{-2/3}\ r_{\rm G}\ (c_0/v_{\rm rel})^{4/3},
 \ee
which is referred to as the Shvartsman radius \citep{Ikhsanov-Finger-2012}. Here $\rho$ is the density and $c_0 = c_{\rm s}(r_{\rm G})$ is the sound speed in the accreting material. $\beta_0 = \beta(r_{\rm G})$, where $\beta = \msE_{\rm th}/\msE_{\rm m}$ and $\msE_{\rm th} = \rho c_{\rm s}^2$ is the thermal pressure in the accretion flow. Studies \citep{Bisnovatyi-Kogan-Ruzmaikin-1974, Bisnovatyi-Kogan-Ruzmaikin-1976, Igumenshchev-etal-2003} have shown that the magnetized spherical flow is decelerated at the distance $R_{\rm sh}$ by its own magnetic field. The material inside Shvartsman radius tends to accumulate in a non-Keplerian slab in which it is confined by the magnetic field of the flow itself. The accretion in the slab proceeds on the magnetic reconnection timescale,  $t_{\rm rec} = r/\eta_{\rm m} v_{\rm A}$, which under conditions of interest (i.e. $v_{\rm A} \leq v_{\rm ff}$ and $\eta_{\rm m} \ll 1$) significantly exceeds the dynamical time. Here $v_{\rm A} = B_{\rm f}/\sqrt{4 \pi \rho}$ is the Alfv\'en velocity in the accreting material and $0 < \eta_{\rm m} \leq 0.1$ is the magnetic reconnection efficiency \citep[][and references  therein]{Somov-2006}. The outer radius of the slab is limited to $r_{\rm out} < R_{\rm sh}$ and its half-thickness can be  approximated by the height of homogeneous atmosphere, $h_{\rm s}(r) = k_{\rm B}\,T(r)\,r^2/m_{\rm p}\,GM_*$, where $m_{\rm p}$ is the proton mass, $k_{\rm B}$ is the Boltzmann constant and $T$ is the gas temperature in the slab. The angular velocity of material in the slab in general case is limited as $0 \leq \Omega_{\rm sl} \leq \Omega_{\rm k} = (GM_*/r^3)^{1/2}$ and the magnetic field scales with radius as $B_{\rm f}(r) \propto r^{-5/4}$ \citep[for discussion see e.g.][]{Bisnovatyi-Kogan-Ruzmaikin-1976}.

A formation of the magnetic slab around a neutron star accreting material from a magnetized wind can be expected if $R_{\rm sh} > r_{\rm A}$, which implies $v_{\rm rel} < v_{\rm ma}$, where
    \be\label{vma}
 v_{\rm ma} \simeq 380\ \beta_0^{-1/5} m^{12/35} \mu_{30}^{-6/35} \dmf_{14}^{3/35} c_6^{2/5}\ {\rm km\,s^{-1}}
 \ee
and $c_6 = c_0/10^6\,{\rm cm\,s^{-1}}$ \citep{Ikhsanov-Finger-2012, Ikhsanov-Beskrovnaya-2012}. Otherwise, the magnetic field of the accretion flow can be neglected.

Finally, for a Keplerian disk to form the circularization radius, $r_{\rm circ} = \xi^2\,\Omega_0^2\,r_{\rm G}^4/GM_{\rm ns}$, should satisfy the condition $r_{\rm circ} > \max\{r_{\rm A}, R_{\rm sh}\}$. Here $\xi$ is a dimensionless parameter accounting for dissipation of angular momentum in the spherical accretion flow \citep{Ruffert-1999} and $\Omega_0 = \Omega(r_{\rm G})$ is the
angular velocity of the material at the Bondi radius. As shown by \citet{Chevalier-1989}, the angular momentum of the material which is initially located close to the neutron star is insignificant. However, the outer mantel material may have a significant angular momentum due to its turbulization by the reverse shock wave. As the turbulence is subsonic the velocity of turbulent motions is $v_{\rm t} = \varepsilon c_0$, where $0 < \varepsilon \leq 1$ and hence, $\Omega_0 = \varepsilon c_0/r_{\rm G}$. The condition $r_{\rm circ} > R_{\rm sh}$ in this case can be expressed as $v_{\rm rel} < v_{\rm cr}$, where
 \be\label{vcr}
 v_{\rm cr} \simeq 4 \times 10^4\ \beta_0^{7/15} \varepsilon^{7/5} \xi_{0.2}^{7/5} \left(\frac{c_0}{10^6\,{\rm cm\,s^{-1}}}\right)^{14/15}\ {\rm cm\,s^{-1}}
 \ee
and $\xi_{0.2} = \xi/0.2$ is normalized according to \citet{Ruffert-1999}.

Thus, the geometry of the fall-back accretion process onto a newly formed neutron star strongly depends on the physical conditions in its environment. It can be approximated by a spherically symmetrical flow if $v_{\rm rel} > v_{\rm ma}$, by a Keplerian disk if $v_{\rm rel} < v_{\rm cr}$ and by a magnetic slab if $v_{\rm cr} < v_{\rm rel} < v_{\rm ma}$. The latter case is discussed in this paper.

 \section{Spin-down torque}\label{torque}

The spin-down torque applied to a neutron star from the magnetic slab can be evaluated as \citep{Ikhsanov-2012}
 \be\label{ksds0}
 |K_{\rm sd}^{\rm (sl)}| = 4 \pi r_{\rm m} h_{\rm s}(r_{\rm m})\,\nu_{\rm m}\ \rho_0\,v_{\phi}(r_{\rm m}).
 \ee
Here $\nu_{\rm m} = k_{\rm m} r_{\rm m} v_{\rm A}(r_{\rm m})$ is the magnetic viscosity, $r_{\rm m}$ is the radius of the magnetosphere of the neutron star and $0 < k_{\rm m} \leq 1$ is the efficiency parameter. $v_{\phi}(r_{\rm m}) = r_{\rm m}\,[\omega_{\rm s} - \Omega_{\rm sl}(r_{\rm m})]$, where $\omega_{\rm s} = 2 \pi/P_{\rm s}$ is the angular velocity of the neutron star. Finally, $\rho_0 = \mu^2\,m_{\rm p}/2 \pi\,k_{\rm B}\,T(r_{\rm m})\,r_{\rm m}^6$ is the gas density at the magnetospheric boundary, which is defined by equating the gas pressure in the slab with the magnetic pressure due to dipole magnetic field of the neutron star.

Combining these parameters and taking into account that the Alfv\'en velocity at the inner radius of the slab is equal to the free-fall velocity \citep{Bisnovatyi-Kogan-Ruzmaikin-1976} one finds
   \be\label{ksdsl1}
|K_{\rm sd}^{\rm (sl)}| = \frac{k_{\rm m}\,\mu^2}{\left(r_{\rm m} r_{\rm cor}\right)^{3/2}}
\left(1 - \frac{\Omega_{\rm sl}(r_{\rm m})}{\omega_{\rm s}}\right),
 \ee
where $r_{\rm cor} = (GM_{\rm ns}/\omega_{\rm s}^2)^{1/3}$ is the corotation radius of the star. Eq.~(\ref{ksdsl1}) represents a generalized form of the spin-down torque applied to a neutron star from the accreting material. It shows that the absolute value of the torque is limited to its conventional value, $\leq |\dmf\,\omega_{\rm s}\,r_{\rm A}^2|$, for $r_{\rm m} \geq r_{\rm A}$, but can be significantly larger if the accreting material approaches the neutron star to a smaller distance than $r_{\rm A}$.

The magnetospheric radius in general case can be evaluated by solving the system
 \be\label{syst1}
 \left\{
 \begin{array}{ll}
 \displaystyle\frac{\mathstrut \mu^2}{2 \pi r_{\rm m}^6} = \rho(r_{\rm m}) c_{\rm s}^2(r_{\rm m}), & \\
  & \\
 \dmf_{\rm in}(r_{\rm m}) = \displaystyle\frac{\mathstrut L_{\rm X} R_{\rm ns}}{GM_{\rm ns}}, & \\
  \end{array}
 \right.
 \ee
where
  \be\label{dmfin-1}
 \dmf_{\rm in}(r_{\rm m}) = 4 \pi r_{\rm m}^{5/4}  \rho_0 (2 GM_{\rm ns})^{1/4} D_{\rm eff}^{1/2}(r_{\rm m})
   \ee
is the rate of plasma diffusion from the slab into the stellar field at the magnetospheric boundary and $D_{\rm eff}$ is the effective diffusion coefficient \citep{Anzer-Boerner-1980, Elsner-Lamb-1984}. The first equation in system~(\ref{syst1}) shows that the pressure of the accreting material at the magnetospheric boundary is equal to the magnetic pressure due to dipole field of the neutron star. The second equation is the continuity equation. It shows that the rate of plasma diffusion into the stellar field at the magnetospheric boundary is equal to the mass accretion rate onto the stellar surface. Solving system~(\ref{syst1}) and taking into account that the effective diffusion coefficient is limited to $D_{\rm eff} \geq D_{\rm B}(r_{\rm m})$ \citep{Elsner-Lamb-1984}, where
 \be
 D_{\rm B}(r_{\rm m}) = \frac{c k_{\rm B} T(r_{\rm m})}{16 e B(r_{\rm m})}
 \ee
is the Bohm diffusion coefficient, one finds that the magnetospheric radius of the neutron star satisfies the condition $r_{\rm m} \geq r_{\rm ma}$, where
  \be\label{rma}
 r_{\rm ma} = \left(\frac{c\,m_{\rm p}^2}{16\,\sqrt{2}\,e\,k_{\rm B}}\right)^{2/13} \frac{\mu^{6/13} (GM_{\rm ns})^{5/13}}{T_0^{2/13} L_{\rm X}^{4/13} R_{\rm ns}^{4/13}}.
 \ee
Here $B(r_{\rm m})$ is the magnetic field strength at the magnetospheric boundary and $T_0$ is the gas temperature at the inner  radius of the slab. The situation $r_{\rm m} = r_{\rm ma}$ is realized if interchange instabilities of the magnetospheric boundary
are suppressed and the accreting material enters the stellar magnetic field due to magnetic reconnection, i.e. in a similar way to the solar wind penetrating into the Earth's magnetosphere \citep{Paschmann-2008}. The spin-down torque applied to the neutron star in this case reaches its maximum possible value (see Eq.~\ref{ksdsl1}).

  \section{Magnetic accretion in 1E1613}\label{current}

Let us consider a situation in which 1E1613 is an isolated neutron star accreting material at the rate $\dmf_*$ from a fossil magnetic slab. The surface magnetic field of the star in this case can be evaluated as $B_* \geq B_{\rm min}$, where
 \be
 B_{\rm min} \simeq 7 \times 10^{11}\ T_6^{1/3} m^{-5/6} R_6^{-7/3} L_{34}^{2/3} \left(\frac{r_{\rm m}}{r_{\rm mag}}\right)^{13/6}\,{\rm G}
 \ee
is the solution of Eq.~(\ref{rma}) for $r_{\rm ma} = r_{\rm mag}$ and $T_6 = T_0/10^6$\,K.

The spin derivative of the pulsar,
 \bdm
|\dot{P}| = \frac{P_{\rm s}^2 |K_{\rm sd}^{\rm (sl)}|}{2 \pi I},
 \edm
in this situation (i.e. $r_{\rm m} = r_{\rm mag}$ and $P_{\rm s} = P_*$) is
     \begin{eqnarray}\label{dotp}
 |\dot{P}_*| & \simeq & 3.3 \times 10^{-7}\,{\rm s\,s^{-1}} \times k_{\rm m}\ \mu_{30}^2\ I_{45}^{-1}\ m^{-1/2}\ \\
       \nonumber
  & \times &  \left(\frac{P_{\rm s}}{P_*}\right) \left(\frac{r_{\rm mag}}{3 \times 10^8\,{\rm cm}}\right)^{-3/2}
 \left(1 - \frac{\Omega_{\rm sl}(r_{\rm ma})}{\omega_{\rm s}}\right),
     \end{eqnarray}
where $I_{45} = I/10^{45}\,{\rm g\,cm^2}$ is the moment of inertia of the neutron star. The condition $|\dot{P}_*| \leq 1.6 \times 10^{-9}\,{\rm s\,s^{-1}}$ implies
 \be
\left|\,1- \frac{\Omega_{\rm sl}(r_{\rm ma})}{\omega_{\rm s}}\right| \leq 0.005\,k_{\rm m}^{-1},
 \ee
which may indicate that the neutron star in the current epoch rotates at almost its maximum possible period $P_{\rm max} = 2 \pi/\Omega_{\rm sl}(r_{\rm ma})$ \citep{Bisnovatyi-Kogan-1991}.

 \section{Possible history}\label{history}

The spin-down time, $\tau \simeq P_{\rm s}/2 \dot{P}$, of a neutron star accreting material from the magnetic slab in the general case can be expressed as
 \be\label{taua}
 \tau_{\rm a} \simeq \frac{P_{\rm s}}{2 \dot{P}_{\rm sl}} =  \frac{I (GM_{\rm ns})^{1/2} r_{\rm ma}^{3/2}}{2 \mu^2},
 \ee
where
 \be
 \dot{P}_{\rm sl} = \frac{|K_{\rm sd}^{\rm (sl)}| P_{\rm s}^2}{2 \pi I} \simeq
 \frac{P_{\rm s}^2 k_{\rm m} \mu^2}{2 \pi I \left(r_{\rm ma} r_{\rm cor}\right)^{3/2}}
 \ee
is the spin-down rate evaluated under the assumption $\Omega_{\rm sl} \ll \omega_{\rm s}$. Putting parameters of 1E1613 derived in the previous section to Eq.~(\ref{taua}) one finds
 \be
 \tau_{\rm a} \simeq 1880\ \mu_{30}^{-17/13} m^{8/13} I_{45} T_6^{3/13} \dmf_{14}^{-6/13}\,{\rm yr}.
 \ee
This shows that a neutron star with the surface magnetic field of $\sim 10^{12}$\,G would have enough time to slow down to a long period if it accretes material from the magnetic slab at an average rate $\sim 10^{14}\,{\rm g\,s^{-1}}$. The spin period which the neutron star is able to reach during this time is limited to $P_{\rm max} \leq 2 \pi/\Omega_{\rm sl}(r_{\rm m})$ and, therefore, is determined by the angular velocity of the material at the inner radius of the slab.

Studies of pulsar population \citep[see e.g.,][and references therein]{Narayan-1987, de-Jager-2008} suggest that the average birth spin period of neutron stars is $P_0 \sim 0.5$\,s. The ejector (spin-powered pulsar) state under these conditions can be avoided if the initial mass-transfer rate in the fall-back accretion process is $\dmf_0 \geq \dmf_{\rm ej}$, where
  \be
  \dmf_{\rm ej} \simeq 8 \times 10^{13}\ f_{\rm m}\,\mu_{30}^2\,v_8^{-1}\,\left(\frac{P_0}{0.5\,{\rm s}}\right)^{-4} {\rm g\,s^{-1}}
  \ee
is the solution of the equation $P_0 = P_{\rm ej}$, and
 \be\label{pej}
 P_{\rm ej} \simeq 0.26\ f_{\rm m}^{1/4}\,\mu_{30}^{1/2}\,\dmf_{15}^{-1/4}\,v_8^{-1/4}\ {\rm s}
 \ee
is the spin period at which the pressure of relativistic wind ejected from the magnetosphere of a newly formed neutron star is equal to the ram pressure of the surrounding material at the Bondi radius \citep[for discussion see e.g.,][]{Ikhsanov-2012}. Here $v_8 = v_{\rm rel}/10^8\,{\rm cm\,s^{-1}}$, $\dmf_{15} = \dmf_0/10^{15}\,{\rm g\,s^{-1}}$ and $f_{\rm m} = 1 + \sin^2{\chi}$, where $\chi$ is the angle between the magnetic and rotational axes of the neutron star \citep{Spitkovsky-2006}.

A higher value of $\dmf_0$ would be required for a newly formed neutron star to avoid the propeller state and to start its spin evolution as an accretor. This situation can be realized if $\dmf_0 \geq \dmf_{\rm pr}$, where
 \be
 \dmf_{\rm pr} \simeq 8 \times 10^{15}\ \mu_{30}^{3/2}\,m^{5/6}\,T_6^{1/2}\,\left(\frac{P_0}{0.5\,{\rm s}}\right)^{-13/6} {\rm g\,s^{-1}}
 \ee
is the solution of the equation $P_0 = P_{\rm pr}$, and
 \be
 P_{\rm pr} \simeq 3.5\ \mu_{30}^{9/13}\,m^{-5/13}\,T_6^{-3/13}\,\dmf_{14}^{-6/13}\,{\rm s}
 \ee
is the spin period defined by equating $r_{\rm ma} = r_{\rm cor}$ \citep[see Eq.~19 in][]{Ikhsanov-2012}.

Thus, one can envisage the following evolution scenario of 1E1613. The neutron star was formed in the core-collapsed supernova explosion with a surface field $B_0 \sim 10^{12}$\,G and a spin period $P_0 \sim 0.5$\,s. It started its spin evolution in the accretor state being surrounded by a magnetic slab formed by the supernova ejecta fall-back. The mass-transfer rate in the slab initially was in excess of $3 \times 10^{15}\,{\rm g\,s^{-1}}$ and had been decreased over 2000\,yr by almost two orders of magnitude. The spin period of the star during this time has been increased to its presently observed value defined by the angular velocity of the material in magnetic slab at the magnetospheric radius. The star in current epoch undergoes accretion from the magnetic slab at the rate $\dmf_*$ and the radius of its magnetosphere is $r_{\rm ma} \sim r_*$. The surface field of the star in the current epoch is about its initial value. The spin-down torque applied to the star from the accreting material is given by Eq.~(\ref{dotp}) in which $\Omega_{\rm sl}(r_{\rm ma}) \sim 0.995\,k_{\rm m}^{-1} \omega_{\rm s}$.

 \section{Discussion}\label{conclusions}

In the frame of the magnetic accretion scenario 1E1613 appears to be a regular isolated neutron star with the conventional values of basic parameters. The extremely long period of the pulsar in this approach is associated with peculiar physical conditions in the stellar environment. Namely, it is assumed that the neutron star is surrounded by a (residual) fossil magnetic slab. This implies that the supernova ejecta was magnetized and the relative velocity of the material moving towards the neutron star after its birth met the condition $v_{\rm cr} < v_{\rm rel} < v_{\rm ma}$ (see Eqs.~\ref{vma} and \ref{vcr}).

Our basic assumption seems rather plausible in the light of current views on the fall-back accretion process. Numerical simulations of supernova explosion \citep[see e.g.,][]{Moiseenko-etal-2006, Endeve-etal-2012} favor a strong magnetization of the ejecta in the vicinity of a newly formed neutron star. Our assumption is also consistent with the results of studies of supernova remnants \citep[see e.g.][and references therein]{Arbutina-etal-2012} which suggest that the magnetic pressure in the environment of the neutron stars is comparable to the thermal gas pressure, i.e. $\beta \sim 1$. The velocity of the material falling-back towards the neutron star after its birth lies in the range $200-2000\,{\rm km\,s^{-1}}$ \citep{Woosley-1988, Chevalier-1989}. This is substantially larger than the typical value of $v_{\rm cr}$ (see Eq.~\ref{vcr}), but is comparable with $v_{\rm ma}$ provided the temperature in the material in the vicinity of the star after its birth is $\sim 10^6 - 10^7$\,K.

Finally, the initial mass-transfer rate in the slab, $\dmf \sim 10^{15} - 10^{16}\,{\rm g\,s^{-1}}$, evaluated in our scenario meets the conditions previously reported by \citet{Chatterjee-etal-2000}. Following their results one can expect that the mass-transfer rate in the slab has been decreased over a time span of $\sim 2000$\,yr by almost 2 orders of magnitude \citep[see also][]{Cannizzo-etal-1990} to its current value $10^{13} - 10^{14}\,{\rm g\,s^{-1}}$. This indicates that a presence of a magnetic slab of the mass $M_{\rm sl} \ga \dmf_{\rm pr} \tau_* \sim 10^{-7}\,{\rm M_{\sun}}$ would be sufficient to supply enough material for powering the X-ray luminosity of the pulsar over 2000\,yr by the accretion process.

 \begin{theacknowledgments}
The research has been partly supported by Israel Ministry of Science, the Program of Presidium of RAS N\,21, Russian Ministry of Science and Education under the grants Nrs.\,8417 and 8394, and NSH-1625.2012.2.
\end{theacknowledgments}

\bibliographystyle{aipproc}   

\end{document}